\documentstyle[preprint,aps,12pt]{revtex}
\begin{document}
\title{\bf Tunneling in $\Lambda$ Decaying Cosmologies
and the Cosmological Constant Problem}
\author{\bf M. A. Jafarizadeh}
\address{Department of Theoretical Physics, Tabriz University, Tabriz  51664, Iran.\\
Institute for Studies in Theoretical Physics and Mathematics, Tehran, 19395-1795, Iran.\\
e-mail: jafarzadeh@ark.tabrizu.ac.ir}
\author{\bf F. Darabi}
\address{Department of Physics, Shahid Beheshti University, Tehran 19839, Iran.\\
Department of Physics, Tarbiyat Moallem University, Tabriz, P.O.Box 51745-406, Iran.\\
e-mail: f-darabi@cc.sbu.ac.ir}
\author{\bf A. Rezaei-Aghdam}
\address{Department of Theoretical Physics, Tabriz University, Tabriz  51664, Iran.\\
Department of Physics, Tarbiyat Moallem University, Tabriz, P.O.Box 51745-406, Iran.\\
e-mail: a-rezaei-a@ark.tabrizu.ac.ir}
\author{\bf A. R. Rastegar}
\address{Department of Physics, Tarbiyat Moallem University, Tabriz, P.O.Box 51745-406, Iran.\\
Institute for Studies in Theoretical Physics and Mathematics, Tehran, 19395-1795, Iran.\\
e-mail: rastgar@ark.tabrizu.ac.ir}

\maketitle
\newpage
\begin{abstract}
The tunneling rate, with exact prefactor, is calculated to first order in $\hbar$
for an empty closed Friedmann-Robertson-Walker (FRW) universe with decaying
cosmological term $\Lambda \sim R^{-m}$ ( $R$ is the scale factor and $m$
is a parameter $0\leq m \leq 2$ ). This model is equivalent to a cosmology with
the equation of state $p_{\chi}=(m/3 -1)\rho_{\chi}$.
The calculations are performed by applying the dilute-instanton
approximation on the corresponding Duru-Kleinert path integral. It is shown that
the highest tunneling rate occurs for $m=2$ corresponding to the cosmic string
matter universe. The obtained most probable
cosmological term, like one obtained by Strominger, accounts
for a possible solution to the cosmological constant problem.
\end{abstract}
\narrowtext
\section{Introduction}

The most accurately measured value of the cosmological constant $\Lambda$ provided
by measurements of the Hubble constant places an upper bound on its present value \cite{S}
$$
|\Lambda|/8\pi G\leq 10^{-29} g/cm^3
$$
According to modern quantum field theory, the structure of vacuum is turned out
to be interrelated with some spontaneous symmetry-breaking effects through the
condensation of quantum (scalar) fields. This phenomenon gives rise to a
non-vanishing vacuum energy density $\rho_{vac}\sim M_p^4$ ( $M_p$ is the planck mass).
The appearance of this characteristic mass scale may have an important effect
on the cosmological constant because it receives potential contributions from
this mass scale due to mass spectrum of corresponding physical fields in quantum
field theory. By taking into account this contribution, an effective (observed)
cosmological constant is defined as the sum of the bare cosmological constant
$\lambda$ and $8\pi G \rho_{vac}$ (\cite{CW, SW}). This type of contribution,
however, gives rise to an immediate difficulty called the cosmological constant problem.
The essence of this problem is that it is very difficult to believe that the bare
cosmological constant $\lambda$ is fine tuned such that the effective cosmological
constant $\Lambda$ satisfies the above inequality after all symmetry-breakings.
There are some possible solutions to this problem rendering $\Lambda$ exactly
or almost exactly vanishing; some outstanding ones are as follows:\\
1) Trying to find some relaxation mechanisms by which $\Lambda$ could relax
to zero or its present small value \cite{1}.\\
2) The Baum-Coleman-Hawking mechanism that wormhole solutions can lead the
cosmological constant $\Lambda$ to become a dynamical variable giving rise to
distribution functions $P(\Lambda)\sim \exp(\frac{3M_p^2}{8 \Lambda})$ and $P(\Lambda)\sim \exp(\exp(\frac{3M_p^2}{8 \Lambda}))$ peaking at $\Lambda=0$ \cite{BCH}.\\
3) A Lorentzian analysis of quantum cosmological tunneling leading to the most
probable value $\Lambda\sim \frac{9M_p^2}{16 R^2}$ at a given scale factor $R$ \cite{ST}.\\
One indirect solution to the cosmological constant problem is also suggested:\\
4) To assume that $\Lambda$ is dynamically evolving and not constant, that is
evolving from very large value to its present small value (\cite{CW, SW}). \\
This last case, although not well stablished, is interesting to the present work.
There are strong observational motivation for considering models in which $\Lambda$
decreases as $\Lambda \sim R^{-m}$ ( $m$ is a parameter).

For $0\leq m<3$  \cite{SW}, the effect of the decaying cosmological constant on the
cosmic microwave background anisotropy is studied and the angular power spectrum
for different values of $m$ and density parameter $\Omega_{m 0}$ is computed.
Models with $\Omega_{m 0}\geq 0.2$ and $m \geq 1.6$ are shown to be in good
agreement with data.

For $m=2$  \cite{CW}, it is shown that in the early universe $\Lambda$ could be
several tens of orders bigger than its present value, but not big enough disturbing
the physics in the radiation-dominant epoch in the standard cosmology. In the
matter-dominant epoch such a varying $\Lambda$ shifts the three space curvature
parameter $k$ by a constant which changes the standard cosmology predictions
reconciling observations with the inflationary scenario. Such a vanishing cosmological
constant also leads to present creation of matter with a rate comparable to that
in the steady-state cosmology. Although the ansatz $\Lambda\sim R^{-2}$ does
not directly solve the  cosmological constant problem but reduces it and the
age problem to one and the same ``reduced problem'': {\em Why our universe can be
so old aged having a radius $R$ much larger than the planck scale}? \cite{CW}

In the present work we advocate the possibility that $\Lambda$ varies as
$R^{-m}(t)$ in favor of the forth solution to the cosmological constant problem.
We approach this problem indirectly in the sense that we try to find a possible
solution to the reduced problem instead of cosmological constant problem. We
shall study the quantum tunneling for an empty closed FRW cosmology with
$\Lambda \sim R^{-m}(t)$ ($0\leq m \leq 2$) as effectively being a cosmological model with
an exotic $\chi$ fluid with the equation of state $p_{\chi}=(m/3 -1)\rho_{\chi}$. Then we calculate the tunneling rate
for this cosmology and show that the maximum tunneling rate corresponds to
$\Lambda\sim R^{-2}$ ($m = 2$) as the most probable cosmological term
as obtained by Strominger \cite{ST}. Therefore, we obtain a reasonable answer to the
reduced problem insisting on the birth from ``nothing'' of our universe through
the tunneling effect consistent with $\Lambda\sim R^{-2}$ simply because it
can lead, through various symmetry-breakings \cite{CW}, to an old aged universe after tunneling.
It is worht emphasizing that {\em some connections between quantum tunneling and
inflation} \cite{Norb} have already been discussed whose relation to this work
may deserve further investigations.

We shall calculate the tunneling rate by applying the dilute-instanton approximation
to first order in $\hbar$ \cite{CCB}, on the corresponding Duru-Kleinert path integral \cite{JDR}.
Its prefactor is calculated by the heat kernel method \cite{L}, using the
shape invariance symmetry \cite{JF}.

This paper is organized as follows:
In section ${\bf 2}$, the Duru-Kleinert path integral formula and Duru-Kleinert
equivalence of corresponding actions is briefly reviewed. In section ${\bf 3}$,
we introduce the cosmological model of a closed FRW universe filled with an exotic
fluid matter. This is effectively an empty closed FRW universe with an
$R$ varying cosmological term. Finally in section ${\bf 4}$, the tunneling
rate for this model is fully calculated to first order in $\hbar$ by applying the dilute-instanton approximation on the corresponding
Duru-Kleinert path integral. The paper is ended with a conclusion.

\section{\bf Duru-Kleinert Path Integral}

$\; \; \; \;$ In this section we briefly review the Duru-Kleinert path integral \cite{DK}.
The fundamental object of path integration is the time displacement
amplitude or propagator of a system, $ (X_b \: t_b\: | \: X_a \: t_a) $.
For a system with a time independent Hamiltonian, the object
$ (X_b \: t_b \: | \: X_a \: t_a) $ supplied by a path integral is the causal
propagator
\begin{equation}
(X_b \: t_b \: | \: X_a \: t_a)=\theta(t_a-t_b)<X_b|\exp(-i\hat{H}(t_b-t_a)/\hbar)|X_a>.
\end{equation}
Fourier transforming the causal propagator in the time variable, we
obtain the fixed energy amplitude
\begin{equation}
(X_b \: | \: X_a \: )_E = \int_{t_a}^\infty dt_b e^{iE(t_b-t_a)/\hbar}
(X_b \: t_b\: | \: X_a \: t_a).
\end{equation}
This amplitude contains as much information on the system as the propagator
$(X_b \: t_b\: | \: X_a \: t_a)$, and its path integral form is as follows:
\begin{equation}
(X_b \: | \: X_a)_E = \int_{t_a}^{\infty} dt_b \int {\cal D}x(t) e^{i{\cal A}_E/\hbar}
\label{a}
\end{equation}
with the action
\begin{equation}
{\cal A}_E = \int_{t_a}^{t_b} dt [\frac{M}{2}\dot{x}^2(t)-V(x(t))+E]
\end{equation}
where $ \dot{x} $ denotes the derivatives with respect to $ t $ .
In \cite{DK} it has been shown that fixed energy amplitude (\ref{a}) is equivalent
to the following fixed energy amplitude,
\begin{equation}
(X_b \: | \: X_a)_E = \int_{0}^{\infty} dS [f_r(x_b)f_l(x_a)\int {\cal D}x(s)
e^{i{\cal A}_{E}^{f}/\hbar}]
\end{equation}
with the action
\begin{equation}
{\cal A}_{E}^{f} = \int_{0}^{S} ds \{ \frac{M}{2f(x(s))}x'^2(s)-f(x(s))
[V(x(s))-E] \}
\label{b}
\end{equation}
where $ f_r $ and $ f_l $ are arbitrary regulating functions such that
$f=f_{l} f_{r}$ and $ x'$ denotes the derivatives with respect to time $s$.
The actions $ {\cal A}_E $ and $ {\cal A}_{E}^{f} $,
both of which lead to the same fixed-energy amplitude $ (X_b \: | \: X_a)_E $ are called
Duru-Kleinert equivalent \footnote{Of course a third action
$ {\cal A}_{E,\varepsilon}^{DK} $ is also Duru-Kleinert equivalent of
$ {\cal A}_E $ and $ {\cal A}_E^f $ but we do not consider it here \cite{DK}.}.

In the following section we shall use this equivalence to calculate the quantum
tunneling rate. For a quantum-mechanical decay of the ground state, the standard
instanton calculation \cite{CCB} yields the transition amplitude
\begin{equation}
<f|i>\equiv \int {\cal D}q \exp\left[-i \int_0^T \![\frac{1}{2} {\dot{x}}^2 -V] dt\right]\simeq e^{-\Gamma T}
\label{fi}
\end{equation}
where $\Gamma$ is the tunneling rate. The essential feature of (\ref{fi})
is that the ground state energy of the corresponding Hamiltonian picks up a small
imaginary part $\Gamma$ signaling the instability. In the instanton calculation
this is taken care by the negative mode in the bounce solution. Note that the basic object
in these calculations is the transition amplitude which plays a key role in the
Duru-Kleinert equivalence.
It is well-known that for a quantum-cosmological tunneling we should impose the ``zero energy '' condition
on the corresponding transition amplitude.
Thus we rewrite the action $ {\cal A}_{E}^{f} $ in a suitable form such that
it describes a system with zero energy; as only in this sense can we describe
a quantum cosmological model with zero energy.
Imposing $ E = 0 $ in (\ref{b}), with a simple manipulation, gives
\begin{equation}
{\cal A}_{E}^{f} = \int_{0}^{1} ds' S f(X(s')) \{ \frac{M}{2[Sf(X(s'))]^2}
\dot{X}^2(s')-V(X(s')) \}
\end{equation}
where $ \dot{X} $ denotes the derivative with respect to new parameter $ s' $ defined by
\begin{equation}
s' = S^{-1} s
\end{equation}
with $S$ as a  dimensionless scale parameter.\\
After a Wick rotation $ s'=-i\tau $, we get the required Euclidean action and
the path integral
\begin{equation}
I_{0}^{f} = \int_{0}^{1} d\tau S f(X(\tau)) \{ \frac{M}{2[Sf(X(\tau))]^2}
\dot{X}^2(\tau)+V(X(\tau)) \},
\label{z}
\end{equation}
\begin{equation}
(X_b \: | \: X_a) = \int_{0}^{\infty} dS [f_r(X_b)f_l(X_a) \int{\cal D}X(\tau)
e^{{-I_{0}^{f}}/\hbar}]
\end{equation}
where $\tau$ is the Euclidean time. The action (\ref{z}) is Duru-Kleinert equivalent
of
\begin{equation}
I_{0} = \int_{\tau_a}^{\tau_b} d\tau [ \frac{M}{2}\dot{X}^2(\tau)+V(X(\tau)) ]
\end{equation}
where $\tau_a$ and $\tau_b$ correspond to $t_a$ and $t_b$ respectively, and $\dot{X}$ denotes the derivative
with respect to Euclidean time $\tau$.

\section{\bf Model}

We shall consider a closed FRW universe filled with an exotic fluid having the
equation of state $p_{\chi}=(m/3 -1)\rho_{\chi}$ with the parameter $m$ restricted to
the range $0 \leq m \leq 2$. Such fractional equation of state is possible since
the exotic matter may have an effective equation of state anywhere in a range
between well established values. For instance, for cosmic strings
$\frac{2}{3} \leq \frac{m}{3} \leq \frac{4}{3}$, and for domain walls
$\frac{1}{3} \leq \frac{m}{3} \leq \frac{4}{3}$, depending on their velocities \cite{KT}.

The system has only one collective coordinate, namely, the scale factor
$R$. Using the usual Robertson-Walker metric we obtain the scalar curvature
\begin{equation}
{\cal R} = 6 \left[\frac{\ddot{R}}{R}+ \frac{1+{\dot{R}}^2}{R^2} \right].
\label{c}
\end{equation}
Substituting (\ref{c}) into the Einstein-Hilbert action plus a matter term indicating
an exotic $\chi$ fluid with the equation of state $p_{\chi}=(m/3 -1)\rho_{\chi}$
leads to the action \footnote{In what follows we shall take units such that $8\pi G=1$.}
\begin{equation}
I=\int_0^1 \! dt \left[-\frac{1}{2}R {\dot{R}}^2+\frac{1}{2}R\left(1-\frac{\rho_{\chi}}{3} R^2 \right) \right]
\label{d}
\end{equation}
with the constraint of Einstein equation
\begin{equation}
{\dot{R}}^2+ \left[1-\frac{\rho_{\chi}}{3} R^2 \right]=0.
\label{e}
\end{equation}
It is easy to show that the equation of state $p_{\chi}=(m/3 -1)\rho_{\chi}$ upon substitution
into the continuity equation $\frac{d\rho_{\chi}}{dR}=-\frac{3}{R}(\rho_{\chi}+p_{\chi})$ leads
to the following behaviour of the energy density in a closed FRW universe \cite{Atk}
\begin{equation}
\rho_{\chi}(R)=\rho_{\chi}(R_0)\left(\frac{R_0}{R}\right)^m
\end{equation}
Now, we may define the cosmological term
$$
\Lambda \equiv \rho_{\chi}(R)
$$
which leads to
\begin{equation}
I=\int_0^1 \! dt \left[-\frac{1}{2}R {\dot{R}}^2+\frac{1}{2}R\left(1-\frac{\Lambda}{3} R^2 \right) \right]
\label{d'}
\end{equation}
and
\begin{equation}
{\dot{R}}^2+ \left[1-\frac{\Lambda}{3} R^2 \right]=0.
\label{e'}
\end{equation}
Using $\Lambda \equiv \rho_{\chi}$, we may have equivalently
\begin{equation}
\Lambda(R)=\Lambda(R_0) \left(\frac{R_0}{R}\right)^{m}
\label{f}
\end{equation}
where $R_0$ is the value of the scale factor at an arbitrary reference time.
It is worth emphasizing that one possible explanation for a small $\Lambda$ term is
to assume that it is dynamically evolving and not constant, that is, as the universe
evolves from an earlier hotter and denser epoch, the effective cosmological term
also evolves and decreases to its present value \cite{Ozer}. There are also strong
observational motivations for considering cosmological models with a decaying $\Lambda$
term instead of a constant one \cite{SW}.

Chen and Wu \cite{CW} have given some interesting arguments in favour of a
cosmological term $\Lambda \sim R^{-2}$ which was phenomenological and did not come
from particle physics first principles. This behaviour could be obtained under
some simple and general assumptions conforming quantum cosmology. From
dimensional considerations one can always write $\Lambda$ as $M_{pl}^4$ times a
dimensionless product of quantities. Supposing that no other parameters are
relevant except the scale factor $R$, the natural ansatz is that $\Lambda$
varies according to a power law in $R$ as \cite{CW}
$$
\Lambda(R) \sim M_{pl}^4 \left(\frac{R_{pl}}{R} \right)^m \:\:\:\:\:(\mbox{with}\:\: \hbar=c=1)
$$
where $M_{pl}$ and $R_{pl}$ are the planck mass and the planck lenght respectively.

Silveira and Wega \cite{SW} have also suggested a class of models in which $\Lambda$
decreases as a power-law dependence on the scale factor $\Lambda \sim R^{-m}$,
where $m$ is a constant $(0\leq m \leq 3)$. Recently they investigated some properties
of flat cosmologies with a cosmololgical term as \cite{SW}
$$
\Lambda=8 \pi G \rho_{vac}=3\alpha R^{-m}
$$
with $\alpha\geq 0$ and $0\leq m<3$. These models are equivalent to standard cosmology
with matter and radiation plus an exotic fluid with the equation of state
$p_{\chi}=(m/3 -1)\rho_{\chi}$. They studied the effect of the decaying $\Lambda$ term
on the cosmic microwave background anisotropy and computed the angular power
spectrum for different values of $m$ and density parameter $\Omega_{m 0}$.

It is to be noted that regarding the
equation of state $p_{\chi}=(m/3 -1)\rho_{\chi}$ with $0\leq m \leq 2$, our model resembles
a negative pressure matter universe violating the strong energy condition.
Dabrowski \cite{Dabr} has already considered
similar problem for oscillating closed Friedmann models with matter source being
domain walls (which scale like $R^{-1}$ ) and negative cosmological constant.
Domain walls are of course an example of matter violating strong energy condition
since for them $m=1$. Cosmic strings on the other hand have $m=2$. Thus, we can
reinterpret them as decaying cosmological terms
\footnote{Private communication with M. P. Dabrowski.}.

Based on these $\Lambda$ decaying models, we were motivated to take the present
model in which a time dependent $\Lambda$ term with a power-law dependence on
the scale factor $R$ is considered.

By introducing a new parameter $\alpha$
restricted to the range $1\leq \alpha <\infty$ we may rewite (\ref{f}) as
\begin{equation}
\Lambda(R)=\Lambda(R_0) \left(\frac{R_0}{R}\right)^{2-\frac{2}{\alpha}},
\label{5}
\end{equation}
The case $m=2$, having some interesting implications
in reconciling observations with the inflationary models \cite{CW}, may be obtained
as $\alpha \rightarrow \infty$. Also, this value for $m$ accounts for an exotic
fluid matter source, namely the cosmic string.
Substituting (\ref{5}) into the action (\ref{d}) leads to
\begin{equation}
I=\int_0^1 \! dt \left[-\frac{1}{2}R {\dot{R}}^2+\frac{1}{2}R\left(1-\left(\frac{R}{R_0}\right) ^{\frac{2}{\alpha}} \right) \right],
\label{g}
\end{equation}
\begin{equation}
{\dot{R}}^2+ \left[1-\left(\frac{R}{R_0}\right) ^{\frac{2}{\alpha}} \right]=0
\end{equation}
where $\Lambda(R_0)=\frac{3}{{R_0}^2}$.
The issue of quantum tunneling for this $\Lambda$ decaying
model may be investigated in two ways: WKB approximation, and dilute-instanton
approximation techniques. In the first one, we may solve the corresponding
Wheeler-DeWitt equation obtaining the tunneling wave functions to calculate
the tunneling probability $\Gamma$, where in the second one we may solve the
Euclidean field equations obtaining instanton solutions to calculate $\Gamma$.
Here, in order to calculate $\Gamma$ we shall follow the second approach.

\section{\bf Tunneling rate}

The Euclidean form of the action (\ref{g}) is not suitable to be used in instanton
calculation techniques. The reason is that the kinetic term is
not in its standard quadratic form. It has been recently shown \cite{JDR} that in
such cosmological model one may use the Duru-Kleinert equivalence to work with
the standard form of the action. Using the same procedure, we find the Duru-Kleinert
equivalent action in the cosmological model here as follows
\begin{equation}
I_0=\int_{\tau_a}^{\tau_b}\! dt \left[\frac{1}{2} {\dot{R}(\tau)}^2+\frac{1}{2}R^2\left(1-\left(\frac{R}{R_0}\right) ^{\frac{2}{\alpha}} \right) \right]
\label{h}
\end{equation}
Now, the Euclidean action (\ref{h}) has the right kinetic term to be used in
instanton calculations. The Euclidean type Hamiltonian corresponding to the
action (\ref{h}) is given by
\begin{equation}
H_{E} = \frac{\dot{R}^2}{2} -\frac{1}{2}R^2 \left[1-\left(\frac{R}{R_0}\right) ^{\frac{2}{\alpha}} \right]
\end{equation}
whose vanishing constraint $H_{E}=0$
\footnote{The constraint $H_{E}=0$ corresponds to Euclidean form of the
Einstein equation.} gives a non-trivial instanton solution
\begin{equation}
R(\tau)= \frac{R_0}{(\cosh(\frac{\tau}{\alpha}))^\alpha}
\label{i}
\end{equation}
corresponding to the potential
\begin{equation}
V(R) = \frac{1}{2}R^2 \left[1-\left(\frac{R}{R_0}\right) ^{\frac{2}{\alpha}} \right]\:\:\:\:\:For \:R\geq0.
\label{j}
\end{equation}
Each solution with $\alpha>0$
describes a particle rolling down from the top of the potential $ -V(R) $
at $ \tau \rightarrow -\infty $ and $ R = 0 $, bouncing back at $ \tau = 0 $ and
$ R = R_0 $ and finally reaching the top of the potential at $ \tau \rightarrow
+\infty $ and $ R = 0 $.\\
The region of the barrier $ 0 < R < R_0 $ is classically forbidden for the zero energy
particle, but quantum mechanically it can tunnel through it with a tunneling
probability which is calculated using the instanton solution (\ref{i}).\\
The quantized FRW universe is mathematically equivalent to this particle, such
that the particle at $ R = 0 $ and $ R = R_0 $ represents ``nothing'' and ``FRW''
universes respectively. Therefore one can find the probability
$$
|<FRW(R_0) \: | \: nothing>|^2 .
$$
The rate of tunneling $ \Gamma $ is calculated through the dilute instanton
approximation to first order in $\hbar$ as \cite{CCB}
\begin{equation}
\Gamma = [\frac{det'(-\partial_{\tau}^2 + V''(R))}{det(-\partial_{\tau}^2 + \omega^2)}]^{-1/2}
e^{\frac{-I_0(R)}{\hbar}} [\frac{I_0(R)}{2\pi\hbar}]^{1/2}
\label{p}
\end{equation}
where $det'$ is the determinant without the zero eigenvalue,\, $ V''(R) $ is the
second derivative of the potential at the instanton solution (\ref{i}),
$ \omega^2 =V''(R)|_{R=0}$ with $\omega^2=1$ for the potential (\ref{j}),
and $ I_0(R) $ is the corresponding Euclidean action evaluated at the instanton solution (\ref{i}).
The determinant in the numerator is defined as
\begin{equation}
det'[-\partial_{\tau}^2 + V''(R)] \equiv \prod_{n=1}^{\infty}|\lambda_n|
\end{equation}
where $ \lambda_n $ are the non-zero eigenvalues of the operator
$ -\partial_{\tau}^2 + V''(R) $.\\
The explicit form of this operator is obtained as
\begin{equation}
O \equiv \alpha^{-2}[-\frac{d^2}{dx^2} - \frac{(\alpha+1)(\alpha+2)}{\cosh^2 x}+\alpha^2]
\label{k}
\end{equation}
where we have used (\ref{i}) and (\ref{j}) with a change of variable $ x = \frac{\tau}{\alpha} $.
Now, in order to find exactly the
eigenvalues and eigenfunctions of the operator (\ref{k}) we assume $\alpha$ to be
positive integer. By relabeling $l=\alpha+1$, the eigenvalue equation
of the operator (\ref{k}) can be written as
\begin{equation}
\Delta_l\psi_l(x)=(E_l-2l+1)\psi_l(x)
\label{y}
\end{equation}
with
\begin{equation}
\Delta_l:=-\frac{d^2}{dx^2}-\frac{l(l+1)}{\cosh^2 x}+l^2
\end{equation}
where the factor $\alpha^{-2}$ is ignored for the moment. The equation (\ref{y}) is a
time independent Schrodinger equation.
Now, by ignoring the constant shift of energy $2l-1$ and by
introducing the following first order differential operators
\begin{equation}
\left\{\begin{array}{ccc}B_l(x): \:=\:\frac{d}{dx}+l\tanh{x} \\
B^\dagger_l(x): \:=\:-\frac{d}{dx}+l\tanh{x},\end{array}
\right.
\end{equation}
the operator $\Delta_l$ can be factorized and
using the shape invariance symmetry we have \cite{JF}
\begin{equation}
\psi_l(x)= \frac{1}{\sqrt{E_l}}B^\dagger_l(x)\psi_{l-1}(x)
\end{equation}
\begin{equation}
\psi_{l-1}(x)= \frac{1}{\sqrt{E_l}}B_l(x)\psi_l(x)
\end{equation}
Therefore, for a given $l$, its first (bounded) excited state can be obtained from
the ground state of $l-1$.
Consequently, the excited state $m$ of a given $l$, that is $\psi_{l,m}$,
can be written as
\begin{equation}
\psi_{l,m}(x)=\sqrt{\frac{2(2m-1)!}{\Pi^m_{j=1} j(2l-j)}}\frac{1}{2^m(m-1)!}
B^\dagger_l(x)B^\dagger_{l-1}(x) \cdots B^\dagger_{m+1}(x) \frac{1}{\cosh^m x},
\end{equation}
with eigenvalues $E_{l,m}=l^2-m^2$.
Also its continuous spectrum consists of
\begin{equation}
\psi_{l,k}=\frac{B_{l}^\dagger(x)}{\sqrt{k^2+l^2}}\frac{B_{l-1}^\dagger(x)}{\sqrt{k^2+(l-1)^2}} \cdots
\frac{B_{1}^\dagger(x)}{\sqrt{k^2+1^2}}\frac{e^{ikx}}{\sqrt{2\pi}},
\label{t}
\end{equation}
with eigenvalues $E_{l,k}=l^2+k^2$ where $\int^{+\infty}_{-\infty}\psi^{*}_{l,k}(x)\psi_{l,k^{\prime}}(x)dx
=\delta(k-k^{\prime})$.
Now, we can calculate the ratio of the determinants as follows.
First we explain very briefly how one can calculate the determinant of an
operator by the heat kernel method \cite{L}. We introduce the generalized
Riemann zeta function of the operator $A$ by
\begin{equation}
\zeta_A(s) = \sum_{m} \frac{1}{|\lambda_m|^s},
\label{m}
\end{equation}
where $ \lambda_m $ are eigenvalues of the operator $A$,\, and the determinant
of the operator $A$ is given by
\begin{equation}
det \, A = e^{-\zeta'_{A}(0)}.
\label{n}
\end{equation}
It is obvious from the equations (\ref{m}) and (\ref{n}) that
for an arbitrary constant $c$
\begin{equation}
det(c A)=c^{\zeta_A(0)}det A .
\label{r}
\end{equation}
On the other hand $ \zeta_A(s) $ is the Mellin transformation of the heat kernel
$ G(x,\, y,\, \tau)$
\footnote{Here $\tau$ is a typical time parameter.}
which satisfies the following heat diffusion equation
\begin{equation}
A \, G(x,\,y,\, \tau) = -\frac{\partial \, G(x,\,y,\, \tau)}{\partial \tau},
\label{o}
\end{equation}
with an initial condition $ G(x,\,y,\,0) = \delta(x - y) $.\, Note that
$ G(x,\,y,\, \tau) $ can be written in terms of its spectrum
\begin{equation}
G(x,\,y,\, \tau) =  \sum_{m} e^{-\lambda_{m}\tau} \psi_{m}^{*}(x) \psi_{m}(y).
\end{equation}
An integral is written for the sum if the spectrum is continuous.
From relations (\ref{n}) and (\ref{o}) it is clear that
\begin{equation}
\zeta_{A}(s) = \frac{1}{\Gamma(s)} \int_{0}^{\infty} d\tau \, \tau^{s-1}
\int_{-\infty}^{+\infty} dx \, G(x,\,x,\, \tau).
\label{w}
\end{equation}
Now, in order to calculate the ratio of the determinants in (\ref{p}), called a
prefactor, we need to find the difference of the functions
$ G(x, y, \tau) $ for two operators $\Delta_l , \Delta_l(0)$, where
\begin{equation}
\Delta_l(0):=-\frac{d^2}{dx^2}+l^2 .
\end{equation}
Considering the fact that $\Delta_l+1-2l$ (or $\Delta_l(0)+1-2l$) has the
same eigen-spaces as $\Delta_l$ (or $\Delta_l(0)$) and the eigen-spectrum
is shifted by $1-2l$, we have
\begin{equation}
G_{\Delta_l(0)+1-2l}(x,y,\tau)=\frac{e^{-(l-1)^2 \tau}}{2\sqrt{\pi\tau}}e^{-\frac{(x-y)^2}{4\tau}}
\end{equation}
\begin{eqnarray}
\nonumber
G_{\Delta_l+1-2l}(x,y,\tau)=\sum^{l-1}_{m=0,m\neq 1}\psi^{*}_{l,m}(x)\psi_{l,m}(y)e^{-|(m-1)(2l-(m+1))|\tau}+\\
 \int^{+\infty}_{-\infty}dke^{-((l-1)^2+k^2)\tau}\psi^{*}_{l,k}(x)\psi_{l,k}(y).
\end{eqnarray}
In order to  calculate the function $\zeta_{\Delta_l+1-2l}$, according
to the relation (\ref{w}) we have to take the trace of heat kernel $G_{\Delta_l+1-2l}(x,y,\tau)$
where we need to integrate over $|\psi_{l,k}|^2$. Using the relation
$\frac{B_l}{\sqrt{E_{l,k}}}\psi_{l,k}(x)=\psi_{l-1,k}(x) $ we have
\begin{eqnarray}
\nonumber
\int^{+\infty}_{-\infty}dx\psi^{*}_{l,k}(x)\psi_{l,k}(x)=-\lim_{x\rightarrow\infty}\frac{1}{\sqrt(E_{l,k})}\psi^{*}_{l,k}(x)\psi_{l-1,k}(x) +\\
\lim_{x\rightarrow -\infty}\frac{1}{\sqrt(E_{l,k})}\psi^{*}_{l,k}(x)\psi_{l-1,k}(x) +\int^{+\infty}_{-\infty}dx\psi^{*}_{l-1,k}(x)\psi_{l-1,k}(x).
\label{u}
\end{eqnarray}
The first and the second terms appearing on the right hand side of the recursion
relation (\ref{u}) are proportional to the asymptotic value of the wave functions at
$ \infty $ and $-\infty $, respectively, where the latter is calculated as
\begin{eqnarray}
\nonumber
\lim_{x\rightarrow \pm\infty} \psi_{m,k}(x)=\frac{-ik\pm m }{\sqrt{k^2+m^2}}\frac{-ik\pm (m-1) }{\sqrt{k^2+(m-1)^2}}\cdots \frac{-ik\pm 1 }{\sqrt{k^2+1}}\frac{\exp(ikx)}{\sqrt{2\pi}}=\\
\frac{1}{\sqrt{2\pi}}\prod^m_{j=1}(\frac{-ik\pm j }{\sqrt{k^2+j^2}})\exp(ikx).
\end{eqnarray}
Substituting these asymptotic behaviours in the recursion
relations between the norms of the wave functions $\psi_{m,k}$
associated with the continuous spectrum (\ref{t}), then using the obtained
recursion relations together with the orthonormality of discrete spectrum
we get the following result for the difference of traces of heat kernels:
\begin{eqnarray}
\nonumber
\int^{+\infty}_{-\infty}dx(G_{\Delta_l+1-2l}(x,x,\tau)-G_{\Delta_l(0)+1-2l}(x,x,\tau))=\;\;\;\;\;\;\;\;\;\;\;\;\;\;\;\;\;\;\;\;\;\;\;\;\;\;\;\;  \\
\nonumber
\sum_{m=0,m \neq 1}^{l-1}\exp(-|(m-1)(2l-(m+1))|\tau)-\frac{1}{\pi}\sum_{m=1}^{l}m(\int^{+\infty}_{-\infty}dk\frac{\exp(-((l-1)^2+k^2)\tau}{(k^2+(l-1)^2)}\\
+((l-1)^2-m^2)\int^{+\infty}_{-\infty}dk\frac{\exp(-((l-1)^2+k^2)}{(k^2+m^2)(k^2+(l-1)^2)})\;\;\;\;\;\;\;\;\;\;\;\;\;\;\;\;\;\;\;\;\;\;\;\;\;\;\;\;\;\;\; .
\label{s}
\end{eqnarray}
Hence, using the Mellin transformation (\ref{w}) and the well-known Feynman integral (A1)
we finally get
\begin{eqnarray}
\nonumber
\zeta_{\Delta_l+1-2l}(s)-\zeta_{\Delta_l(0)+1-2l}(s)=\;\;\;\;\;\;\;\;\;\;\;\;\;\;\;\;\;\;\;\;\;\;\;\;\;\;\;\;\;\;\;\;\;\;\;\;\;\;\;\;\;\;\;\;\;\;\; \\
\nonumber
\sum_{m=0,m \neq 1}^{l-1}(|(m-1)(2l-(m+1))|)^{-s}-\frac{1}{2\pi}l(l+1)(l-1)^{-(2s+1)}\beta(s+\frac{1}{2},\frac{1}{2})\;\;\;\;\;\;\;\;\;\;\;\;\;\;\;\;\;\;\;\;\; \\
\nonumber
-\frac{1}{\sqrt{\pi}}\frac{\Gamma(s+\frac{3}{2})}{\Gamma(s+2)}\sum_{m=1}^{l-1}m(l-1)^{-(2s+3)}((l-1)^2-m^2)_2F_1(s+\frac{3}{2},1,s+2,1-\frac{m^2}{(l-1)^2})\\
\nonumber
\hspace{-80mm}-\frac{1}{\sqrt{\pi}}\frac{\Gamma(s+\frac{3}{2})}{\Gamma(s+2)}l^{-2(s+1)(1-2l)}_2F_1(s+\frac{3}{2},s+1,s+2,1-\frac{(l-1)^2}{l^2})\;\;\;\;\;\; \\
\label{v}
\end{eqnarray}
where $\beta$ is the $beta$ function.\\
For $s=0$, we obtain
\begin{equation}
\zeta_{\Delta_l+1-2l}(s)-\zeta_{\Delta_l(0)+1-2l}(s)|_{s=0}=-1.
\label{q}
\end{equation}
This means that the operators $ \Delta_l+1-2l $  and $ \Delta_l(0)+1-2l $
have the same number of eigen-spaces (even though  for both of them this
number is infinite) since from the definition of Riemann's zeta function, it is
obvious that its value at $s=0$ can be interpreted as  the number of eigen
spaces of the corresponding operator. The appearance of $-1$ on the right hand
side of the relation is due to the ignorance of the eigen-functions associated
with the zero eigen value of the operator $ \Delta_l+1-2l $. Therefore, its
number of eigen-states is the same as that of the operator $ \Delta_l(0)+1-2l $.
In order to calculate the ratio of the determinant of the operators
$ \Delta_l+1-2l $  and $ \Delta_l(0)+1-2l $ we need to know the derivative
of their associated zeta functions at $s=0$. Hence differentiating both sides of the
relation (\ref{q}) with respect to $s$ and evaluating such integrals as (A2,A3)
we get
\begin{equation}
\zeta^{\prime}_{\Delta_l+1-2l}(s)-\zeta^{\prime}_{\Delta_l(0)+1-2l}(s)|_{s=0}=\log (\frac{2(2l-1)!}{((l-2)!)^2}.
\label{2}
\end{equation}
Therefore, according to the relations (\ref{p}), (\ref{n}), (\ref{r}), (\ref{q}) and (\ref{2}) the prefactor
associated with the potential (\ref{j}), is calculated as
\begin{equation}
\frac{det(\frac{1}{(l-1)^2}(-\frac{d^2}{dx^2}-\frac{l(l+1)}{\cosh^2x}+(l-1)^2))}
{det(\frac{1}{(l-1)^2}(-\frac{d^2}{dx^2}+(l-1)^2))}=\frac{((l-1)!)^2}{2(2l-1)!}.
\end{equation}
Finally, the decay rate of metastable state of this potential is calculated as
\begin{equation}
\Gamma=\frac{1}{\sqrt{\pi\hbar}}R_0 2^{l-1}\exp(-\frac{(l-1)R_0^2 \beta(\frac{3}{2},l-1)}{\hbar})+O(\hbar)
\label{1}
\end{equation}
where we have used the value of $I_0$ at the instanton solution (\ref{i}):
\begin{equation}
I_0=\int_{-\infty}^{\infty}R_0^2 \frac{(\sinh(\frac{\tau}{\alpha}))^2}{(\cosh(\frac{\tau}{\alpha}))^{2(1+\alpha)}}d\tau=\alpha R_0^2 \beta(\frac{3}{2},\alpha).
\end{equation}
\newpage
\section{\bf Conclusion}

In this paper we have calculated the tunneling rate, with exact prefactor, to
first order in $\hbar$ from ``nothing'' to a closed FRW universe with decaying
$\Lambda$. The tunneling rate (\ref{1}) increases
for higher values of the positive integer $\alpha$ (or $l$) such that for
$\alpha \rightarrow\infty$ the most tunneling rate corresponds to the
most probable  cosmological term as $\Lambda\sim R^{-2}$.
This decaying cosmological term may have its origin in the cosmic string as the exotic
matter with the effective equation of state $p_{\chi}=-\frac{1}{3}\rho_{\chi}$.
It is probable that the universe could have tunneled with highest
probability from nothing to an empty closed FRW cosmology with a typical planck
size $R_p$ and a large cosmological term $\Lambda\sim R^{-2}_p$. Then, after
tunneling, $\Lambda$ may have evolved and decreased to its present small value as the
universe has expanded classically. As is discussed in \cite{CW} it does not
directly solve the cosmological constant problem but reduces it with the age
problem to one and the same problem: {\em Why our universe could have escaped
the death at the planck size?}

One possible solution to this problem is that the value of $\Lambda$ after tunneling
might be large enough to derive various symmetry-breakings necessary to the
appearance of a universe which has evolved to the present universe with an small
cosmological constant. This may be consistent with an inflationary model in which
an extraordinarily brief period of rapid expansion occurs where the universe is
about the planck size after quantum tunneling. We remark that this $\Lambda$ decaying
model has some advantages in that alleviates some problems in reconciling
observations with the inflationary scenario; in particular it leads to creation of
matter \cite{CW}.
It is interesting to note that
the tunneling rate in this model is the same one obtained for a closed FRW cosmology
with perfect fluid violating the strong energy condition with the equation of
state $p=(\frac{m}{3}-1)\rho$ \cite{JDRR} such that the most probable
cosmological term corresponding to $m=2$ is equivalent to the least violation of the strong
energy condition. This may account for another possible solution to the
reduced problem from the point of view of {\em energy conditions}. In other words,
one may think that our universe could have escaped the death at the planck size
because the violation of energy conditions is minimized right after quantum tunneling.
\newpage
\section*{Acknowledgment}

We would like to thank M. P. Dabrowski and J. Norbury for useful comments and
kindly remarking us their previous works related to the present one.

\appendix
\section*{A}
\begin{eqnarray}
\frac{1}{D_1^{a_1}D_2^{a_2} \cdots D_n^{a_n}}=\frac{\Gamma(a_1+a_2+ \cdots +a_n)}
{\alpha(a_1)\Gamma(a_2) \cdots \Gamma(a_n)}\int dt_1dt_2 \cdots dt_n\frac
{\delta(1-t_1-t_2 \cdots -t_n)t_1^{a_1-1}t_2^{a_2-1} \cdots t_n^{a_n-1}}
{(t_1D_1+t_2D_2+ \cdots +t_nD_n)^{a_1+a_2+ \cdots +a_n}}.
\end{eqnarray}
\begin{equation}
\int^1_0dt(((l-1)^2-m^2)t+m^2)^{-\frac{3}{2}}\log t=\frac{4}{m((l-1)^2-m^2)}(\log 2+\log m-2\log (l-1+m)),
\end{equation}
\begin{eqnarray}
\nonumber
\int^1_0dt(((l-1)^2-m^2)t+m^2)^{-\frac{3}{2}}\log (((l-1)^2-m^2)t+m^2)\log t=\\
\frac{4}{m((l-1)^2-m^2)}(\log m\frac{m}{l-1}\log(l-1)+\frac{(l-1-m)}{l-1}) m-2\log (l-1+m)),&
\end{eqnarray}

\end{document}